\documentclass[reprint,amsmath,amssymb, prd]{revtex4-1}

\usepackage{graphicx}
\usepackage{dcolumn}

\usepackage{epstopdf}
\usepackage{parskip}
\usepackage{float}
\usepackage{booktabs}
\usepackage{rotating}
\usepackage{textcomp}
\usepackage{subfigure}
\usepackage{url}
\usepackage[margin=10pt,font=small,labelfont=bf, labelsep=endash]{caption}
\usepackage[braket]{physymb}
\usepackage{young}
\usepackage{slashed}
\usepackage{bbold}
\usepackage{tensor}
\usepackage{epstopdf}

\newcommand{\p}{\partial}
\newcommand{\eq}[1]{\begin{align*}#1\end{align*}}
\newcommand{\Eq}[1]{\begin{align}#1\end{align}}

\begin{document}

\preprint{APS/123-QED}

\title{Chiral squaring and KLT relations}

\author{Anders Schreiber}
 \email{wsl980@ku.dk}
\affiliation{%
 Niels Bohr International Academy, Niels Bohr Institute,\\
 University of Copenhagen, Blegdamsvej 17, DK 2100, Copenhagen, Denmark
}%

\date{\today}

\begin{abstract}
We demonstrate that amplitudes based on matter supermultiplets can be combined to provide amplitudes of vector supermultiplets by means of KLT relations. In practice we do this by developing a procedure for removing supersymmetry supercharges from super Yang-Mills theory and supergravity supermultiplets, reducing them to vector and chiral supermultiplets respectively. This way, we reduce the super KLT relations to chiral KLT relations making chiral squaring of amplitudes manifest. We study these chiral KLT relations, discussing permutation symmetry and vanishing relations. Finally some explicit calculations are done to show how the relations work in detail. 
\end{abstract}

\maketitle


\section{Introduction}

In the quest to understand the quantum theory of gravity, we often turn to string theory as quantized modes on the bosonic/super string produces graviton state. To say something relevant about observables in quantum gravity, we need to calculate scattering amplitudes. It has, at least to tree level, become apparent how to calculate gravity amplitudes, as we can make use of some tricks from string theory. The famous Kawai-Lewellen-Tye (KLT) relations \cite{klt} are amplitude relations from string theory relating closed string amplitudes to open string amplitudes. These relations are applicable only to tree level amplitudes, so this is what we shall restrict ourselves to for the rest of this paper. From the stringy KLT relations, we can consider the $\alpha' \rightarrow 0$ limit, where one take strings to be point-like objects. This the field theory limit of the stringy KLT relations. In this limit one obtains amplitudes for gravitons from the closed string amplitudes, while obtaining gauge theory amplitudes from the open string amplitudes. This gives us a tool for computing gravity amplitudes without a reference to the Einstein-Hlibert action and problems associated with quantizing this action. The KLT relations allows for gravity amplitudes to factorize into two copies of gauge theory amplitudes, which is the foundation for the famous gauge/gravity relation. For various applications, see \cite{zviliving, johansson1, johansson2, johansson3, johansson4}. The field theory map for KLT relations can be written in the form \cite{kltfieldtheory, kltkernel}
\eq{
M_n = &\sum_{\gamma, \beta \in S_{n-3} }  A_n (n-1, n, \gamma, 1)\\
&\times \mathcal{S} [\gamma | \beta]_{p_1} \times \tilde{A}_n (1, \beta, n-1, n),
}
where $M_n$ is an $n$-point gravity amplitude, while $A_n$ and and $\tilde{A}_n$ are $n$-point Yang-Mills (YM) amplitudes. The object $\mathcal{S}[\gamma| \beta]_{p_1}$ is the $\mathcal{S}$-kernel which glue together the two YM amplitudes \cite{kltfieldtheory}. It serves the purpose of removing double-poles of the amplitude product, while also ensuring the full symmetry of the gravity amplitude. The YM amplitudes are color-ordered, so the full symmetry of the gravity amplitude is non-trivial, but is made manifest by properties of the $\mathcal{S}$-kernel. 

The choice of the $\mathcal{S}$-kernel is not unique for the KLT relations in string theory \cite{klt}. There is a whole family of these that 'glues' together the YM amplitudes correctly. However, this implies relations for the YM amplitudes. The BCJ relations are these relations and were first conjectured in \cite{bcj} and later shown to follow from monodromy in string theory \cite{monodromy1, monodromy2}. The $\mathcal{S}$-kernel of the KLT map can be seen as a generator of BCJ relations \cite{kltfieldtheory, kltkernel}.

Since KLT relations hold for superstrings as well as bosonic strings, it is also possible to relate supergravity (SUGRA) amplitudes to super-Yang-Mills (SYM) amplitudes. These super KLT relations will be the main tool used in this paper. More generally it has been shown \cite{fullsuperkltproof} that the super KLT relations also hold for superamplitudes \cite{elvang} in $\mathcal{N}=8$ SUGRA and $\mathcal{N}=4$ SYM theory. The fact that the maximally supersymmetric KLT relations are manifest means that we can remove supersymmetry systematically and get KLT relations for theories with less supersymmetry. This was thoroughly explored in \cite{kltmap}.

Here we ask ourselves the question of how matter multiplets can be used to construct generalized KLT relations. This is not \emph{a priori} obvious, but we can make effcient use of known results for theories without matter multiplets to derive the correct relations.

To this end, we develop a new procedure for removing supersymmetry supercharges from maximally supersymmetric multiplets in the KLT relations and hence produce new relations which we call \emph{chiral KLT relations}. As we shall see, the $\mathcal{N}=8$ SUGRA multiplet can be reduced to a vector multiplet while the $\mathcal{N}=4$ SYM multiplets will simultaneously be reduced to  chiral multiplets. The KLT relations thus 'square' the chiral amplitudes to give vector amplitudes. We shall refer to this as \emph{chiral squaring}.

This paper is oranized as follows. We start by introducing the on-shell superspace formalism in Section \ref{sec:onshell}. In Section \ref{sec:klt} we give a review of KLT and super KLT relations. A new procedure for reducing supersymmetry is discussed in Section \ref{sec:susy}. Chiral KLT relations are derived and discussed in Section \ref{sec:chiralklt}. Finally we do some explicit validity checks of the chiral KLT relations in Section \ref{sec:explicit}.


\section{On-shell superspace \label{sec:onshell}}

An important tool that we will use throughout this paper is the on-shell superspace \cite{elvang}. In this formalism we can put all the information of a supermultiplet into a compact superfield description of the multiplet. The field content of $\mathcal{N}=4$ SYM theory is a gluon with helicity $h = +1$: $g_+$, four gluinos with helicity $h=+ \frac{1}{2}$: $f_+^a$, six scalars $s^{ab}$, four gluinos with helicity $h = - \frac{1}{2}$: $f_-^{abc}$ and one gluon with helicity $h = -1$: $g_-^{1234}$. Indices $a,b,c = 1,2,3,4$ are $SU(4)_R$ indices associated with the supercharges $Q^a$. We can write down the field content as a superfield: a polynomial expansion of fields in the multiplet in terms of on-shell Grassmann variables \cite{symsuperfield}
\Eq{  \label{eq:n=4superfield}
\begin{split}
\Phi^{\mathcal{N}=4} &= g_+ + \eta_a f^a_+ + \frac{1}{2!} \eta_a \eta_b s^{ab} \\
&+ \frac{1}{3!} \eta_a \eta_b \eta_c f_-^{abc} + \eta_1 \eta_2 \eta_3 \eta_4 g_-^{1234},
\end{split}
}
where the Grassmann variables $\eta_a$ are labeled with the $SU(4)_R$ indices. The supercharges take the form \cite{elvang}
\Eq{ \label{eq:onshellsupercharges}
\tilde{Q}^{\dot{\alpha}}_a = \sum_{i= 1}^n \ket{i}^{\dot{\alpha}} \eta_a, \quad Q_{\alpha}^a = \sum_{i=1}^n |i]_\alpha \frac{\p}{\p \eta_a}.
}
We can reduce the amount of supercharges by setting $\eta$'s to zero or integrating them out. We can extract specific multiplet fields in equation \eqref{eq:n=4superfield} by applying derivatives with respect to the Grassmann variables and setting the remaining variables to zero. The specific prescription for this can be found in Table \ref{table:n=4superfield}.
\begin{table}[t!] 
\centering \def\arraystretch{1.5}
\begin{tabular}{r | c | c | c | c | c }
$\Phi^{\mathcal{N}=4}(p_i)$: & $g_+ (p_i)$ & $f^a_+ (p_i)$ & $s^{ab}(p_i)$ & $f^{abc}_-(p_i)$ & $g_-^{1234}(p_i)$ \\
 \hline
Operator: & 1 & $\p^a_i $ & $\p^a_i \p^b_i$ & $\p^a_i \p^b_i \p^c_i$ & $\p^1_i \p^2_i \p^3_i \p^4_i$  
\end{tabular}
\captionsetup{justification=raggedright, margin = 20pt} 
\caption{This table gives an overview of how to extract specific fields from the superfield \eqref{eq:n=4superfield}. Here $\p_i^a = \left. \frac{\p}{\p \eta_{a}} \right|_{p = p_i}$. After using the desired operator to pick out a field, set all remaining $\eta$'s to zero.}
\label{table:n=4superfield}
\end{table}
Another important item in the on-shell superspace toolbox is the superamplitude: a scattering amplitude where external legs are in the superfield representation
\Eq{
\mathcal{A}_n^{\mathcal{N}=4} (\Phi_1, \Phi_2, \ldots, \Phi_n).
}
Working with superamplitudes rather than regular amplitudes, we can work out amplitude properties for general field and helicity configurations. To get an amplitude from the superamplitude, we simply use the prescription in Table \ref{table:n=4superfield} on each leg to pick out the desired external data.

Knowing how superamplitudes work and how to deduce amplitudes from these superamplitudes, we will now introduce an expansion of superamplitudes into MHV amplitudes, NMHV amplitudes and so on. We do this by making a classification of amplitudes akin to those of purely gluonic amplitudes \cite{elvang}. An MHV amplitude (for $n$ external gluons) has two negative helicity and $(n-2)$ positive helicity gluons (independent of leg ordering). In analogy we define the MHV part of the superamplitude to be proportional to $(\eta)^8$ (since a minus helicity gluon has $(\eta)^4$ in front of it in the superfield description). For an NMHV amplitude we have three negative helicity gluons, so NMHV part of the superamplitude is proportional to $(\eta)^{12}$  and so on. Henceforth we write \cite{elvang, elvang2}
\Eq{ \label{eq:N=4SYMsuperamplitudeexpansion}
\begin{split}
\mathcal{A}_n^{\mathcal{N}=4} &= \sum A_n^\text{MHV} (\eta)^8 + \sum A_n^\text{NMHV} (\eta)^{12} \\
&+ \cdots + \sum A_n^{\overline{\text{MHV}}} (\eta)^{4n-8}.
\end{split}
}
The numbers in front of the Grassmann variables are component amplitudes. We can extract a component amplitudes $A_n^{\text{N}^k\text{MHV}}$ amplitude using $4k + 8$ differential operators and setting the remaining $\eta$'s to zero in equation \eqref{eq:N=4SYMsuperamplitudeexpansion}. We note that each $SU(4)_R$ indices on the $\eta_a$'s appear the same number of time in each of the monomials. This is to ensure $SU(4)_R$ invariance of the superamplitude.


\section{KLT relations \label{sec:klt}}

In this section, we give a review of the KLT relations. Originally the KLT relations were relations between string theory amplitudes \cite{klt}. However, we are interested in the field theory version of the KLT relations \cite{kltfieldtheory}
\Eq{ \label{eq:klt}
\begin{split}
M_n = &\sum_{\gamma, \beta \in S_{n-3}}  A_n (n-1, n, \gamma, 1) \\
&\times \mathcal{S} [\gamma | \beta]_{p_1} \times \tilde{A}_n (1, \beta, n-1, n),
\end{split}
}
where $M_n$ is an $n$-point gravity amplitude and $\{A_n, \tilde{A}_n \}$ are $n$-point YM amplitudes . We sum over two sets of $(n-3)$ permutations $\beta$ and $\gamma$. The momentum kernel $\mathcal{S}$ 'glues' the gluonic amplitudes $A_n $ and $\tilde{A}_n$ together, removing any double poles and ensuring full symmetry of the graviton amplitude. Explicitly, $\mathcal{S}$ has the form \cite{kltfieldtheory, kltkernel}
\Eq{ \label{eq:s-kernel}
\begin{split}
&\mathcal{S} [i_1, \ldots, i_m| j_1, \ldots, j_m]_{p_1} \\
&\equiv \prod_{t=1}^m \left( s_{i_t, 1} + \sum_{q>t}^m \theta(i_t, i_q) s_{i_t, i_q} \right),
\end{split}
}
where $\theta(i_t, i_q)$ equals 1 if the ordering of the legs $i_t$ and $i_q$ are opposite in the sets $\{ i_1, \ldots , i_k \}$ and $\{j_1, \ldots, j_k \}$ and 0 if the ordering is the same. 

KLT relations can also be formulated in terms of superamplitudes for supersymmetric versions of gravity and YM theory. We already introduced the superfield formalism for $\mathcal{N}=4$ SYM theory in equation \eqref{eq:n=4superfield}. This formalism works in the same way for $\mathcal{N}_G=8$ SUGRA superamplitudes.The $\mathcal{N}_G=8$ SUGRA multiplet consists of one graviton $h_\pm$, 8 gravitinos $\psi_\pm$, 28 graviphotons $\nu_\pm$, 56 graviphotinos $\chi_\pm$ and 70 real scalars $\phi$. Like for the $\mathcal{N}=4$ SYM multiplet, the $\mathcal{N}=8$ SUGRA multiplet fits naturally into a superfield 
\Eq{ \label{eq:n=8superfield}
\begin{split}
\Phi^{\mathcal{N}_G = 8} &= h_+ + \eta_A\psi^A_+ + \frac{1}{2!} \eta_A \eta_B \nu^{AB}_+ \\
&+ \frac{1}{3!} \eta_A \eta_B \eta_C \chi^{ABC}_+ + \frac{1}{4!} \eta_A \eta_B \eta_C \eta_D \phi^{ABCD} \\
&+ \frac{1}{5!} \eta_A \eta_B \eta_C \eta_D \eta_E \chi_-^{ABCDE} \\
&+ \frac{1}{6!} \eta_A \eta_B \eta_C \eta_D \eta_E \eta_F \nu_-^{ABCDEF}  \\
&+ \frac{1}{7!} \eta_A \eta_B \eta_C \eta_D \eta_E \eta_F \eta_G \psi_-^{ABCDEFG} \\
&+ \eta_1 \eta_2 \eta_3 \eta_4 \eta_5 \eta_6 \eta_7 \eta_8 h_-^{12345678}.
\end{split}
}
where $A, B, \ldots = 1,\ldots, 8$ are the $SU(8)_R$ indices associated with the $R$-symmetry of the $\mathcal{N}_G = 8$ superspace. We are now in the position to introduce the superamplitude for $\mathcal{N}_G = 8$ SUGRA
\Eq{ \label{eq:N=8superamplitude}
\mathcal{M}_n^{\mathcal{N}_G = 8} (\Phi_1, \Phi_2, \ldots, \Phi_n).
}
Like for SYM superamplitudes an expansion like \eqref{eq:N=4SYMsuperamplitudeexpansion} also exists for \eqref{eq:N=8superamplitude}. KLT relations can be written in terms of superamplitudes between $\mathcal{N}_G =8$ SUGRA and two copies of $\mathcal{N}=4$ SYM theory
\Eq{ \label{eq:N=8superklt}
\begin{split}
\mathcal{M}_n^{\mathcal{N}_G = 8} = &\sum_{\gamma, \beta \in S_{n-3}} \mathcal{A}_n^{\mathcal{N} = 4}  (n-1 , n, \gamma, 1)  \\
&\times \mathcal{S}[\gamma| \beta]_{p_1} \times \tilde{\mathcal{A}}_n^{\tilde{\mathcal{N}} = 4} ( 1, \beta, n-1, n).
\end{split}
}
where $\gamma, \beta$ are permutations over legs $2, \ldots, n-2$ and the $\mathcal{S}$-kernel, $\mathcal{S}[\gamma| \beta]_{p_1}$, is defined in equation \eqref{eq:s-kernel}. Equation \eqref{eq:N=8superklt} can be proven using super BCFW recursion relations \cite{fullsuperkltproof}. The superamplitude expansions correctly yield all the correct component relations when the $\eta$'s on the supergravity side are correctly identified with the unions of $\eta$'s of the two SYM multiplets.

The relation \eqref{eq:N=8superklt} contains all the the maximal amount of information, as it can be reduced to KLT relations for less supersymmetric multiplets. Since $SU(8)_R \supset SU(4)_R \otimes SU(4)_R $ there is a perfect matching between $SU(4)_R$ indices 1,2,3,4 of $\mathcal{N} = 4$ and the $SU(4)_R$ indices 5,6,7,8 of $\tilde{\mathcal{N}} = 4$  for the amplitudes $\mathcal{A}_n^{\mathcal{N}= 4}$ and $\tilde{\mathcal{A}}_n^{\tilde{\mathcal{N}} = 4}$. The product of these superamplitudes match exactly with the $\mathcal{N}_G = 8$ $SU(8)_R$ indices $1,\ldots,8$. Hence, on the righthand side of equation \eqref{eq:N=8superklt} we get strings of $\eta_{i, a}$'s and $\eta_{i,b}$'s where $i=1,\ldots, n$, $a= 1,2,3,4$ and $b = 5,6,7,8$. These are matched to the lefthand side with strings of $\eta_{i,A}$'s with $A = 1,\ldots, 8$. Henceforth we can pick out the appropriate coefficients of $\eta$-strings on the left- and righthand side of \eqref{eq:N=8superklt} and get KLT relations for the component amplitudes.

Supercharges can be removed by integrating out $\eta$'s or setting $\eta$'s to zero (see equation \eqref{eq:onshellsupercharges}) in superfields. However this leads to a problem as superfields are no longer CPT self conjugate after removing $\eta$'s. For $\mathcal{N}=4$ SYM theory, in order to ensure CPT symmetry of the multiplet field content, we must construct two superfields that are related by CPT conjugation. One superfield is constructed by setting $\eta$'s to zero (we call this the $\Phi$-superfield) while in the other we integrate the same $\eta$'s out (we call this the $\Psi$-superfield). This leads to the $\Phi$-$\Psi$ formalism \cite{kltmap}. A similar procedure is used for supersymmetry reduction of $\mathcal{N}_G=8$ SUGRA superfields.

Applying this procedure at the KLT relation level, we can get KLT relations for $\mathcal{N}_G < 8$ for the SUGRA superamplitude and $\mathcal{N}<4$ for SYM superamplitude
\begin{widetext}
\Eq{ \label{eq:N<8superklt}
\begin{split}
&\sum_{\gamma, \beta \in S_{n-3}}\mathcal{A}_{n, i_1, \ldots, i_m}^{\mathcal{N} \leq 4}  (n-1 , n, \gamma, 1) \mathcal{S}[\gamma| \beta]_{p_1}   \tilde{\mathcal{A}}_{n, \tilde{i}_1, \ldots, \tilde{i}_{\tilde{m}}}^{\tilde{\mathcal{N}} \leq 4} (1, \beta, n-1, n).\\
&= \sum_{\gamma, \beta \in S_{n-3}}  \left[ \int \prod_{\tilde{a}_1 = \tilde{\mathcal{N}}+1}^4 d \eta_{\tilde{i}_1, \tilde{a}_1} \cdots \prod_{\tilde{a}_{\tilde{m}} = \tilde{ \mathcal{N}} + 1}^4 d\eta_{\tilde{i}_{\tilde{m}}, \tilde{a}_{\tilde{m}} }  \mathcal{A}^{\mathcal{N}=4}_{n} (n-1 , n, \gamma, 1)  \right]_{\eta_{\tilde{\mathcal{N}}+1}, \ldots, \eta_4  \rightarrow 0}\\
&\times \mathcal{S}[\gamma| \beta]_{p_1} \times \left[ \int \prod_{a_1 = \mathcal{N}+5}^8 d \eta_{i_1, a_1} \cdots \prod_{a_m = \mathcal{N} + 5}^8 d\eta_{i_m, a_m}  \tilde{\mathcal{A}}^{\tilde{\mathcal{N}}=4}_{n}(1, \beta, n-1, n)  \right]_{\eta_{\mathcal{N}+5}, \ldots, \eta_8 \rightarrow 0} \\
&= \left[ \int \prod_{\tilde{a}_1 = \tilde{\mathcal{N}}+1}^4 d \eta_{\tilde{i}_1, \tilde{a}_1} \cdots \prod_{\tilde{a}_{\tilde{m}} = \tilde{ \mathcal{N}} + 1}^4 d\eta_{\tilde{i}_{\tilde{m}}, \tilde{a}_{\tilde{m}} } \prod_{a_1 = \mathcal{N}+5}^8 d \eta_{i_1, a_1} \cdots \prod_{a_m = \mathcal{N} + 5}^8 d\eta_{i_m, a_m}  \mathcal{M}_n^{\mathcal{N}_G = 8}  \right]_{\substack{\eta_{\tilde{\mathcal{N}}+1}, \ldots, \eta_4  \rightarrow 0 \\ \eta_{\mathcal{N}+5}, \ldots, \eta_8 \rightarrow 0 }} \\
&\equiv \mathcal{M}^{\mathcal{N}_G \leq 8}_{n, (\tilde{i}_1, \ldots, \tilde{i}_{\tilde{m}}); (i_1, \ldots, i_m)},
\end{split}
}
\end{widetext}
where subscripts $(\tilde{i}_1, \ldots, \tilde{i}_{\tilde{m}})$ and $(i_1, \ldots, i_m)$ label the external legs in the $\Psi$ and $\tilde{\Psi}$ representations respectively, with $m \leq n$ and $\tilde{m} \leq n$. See \cite{kltmap} for a more thorough discussion of KLT relations for SUGRA and SYM superamplitudes.


\section{A new procedure for reducing supersymmetry  \label{sec:susy}}

In the usual procedure to remove supersymmetry from multiplets, we either integrate out or set $\eta$'s to zero. In the $\Phi$-$\Psi$ formalism, described briefly in the previous section \cite{kltmap}, one starts with the superfield \eqref{eq:n=4superfield} and construct the $\mathcal{N}=3$ SYM superfields by removing $\eta_4$. The $\Phi$ is constructed by setting $\eta_4$ to zero in \eqref{eq:n=4superfield}
\Eq{ 
\begin{split}
\Phi^{\mathcal{N} =3} &= \left. \Phi^{\mathcal{N}=4} \right|_{\eta_4 \rightarrow 0} \\
&=  g_+  + \eta_1 f^1_+ + \eta_2 f^2_+ + \eta_3 f^3_+   +  \eta_1 \eta_2 s^{12} \\
&+  \eta_1 \eta_3 s^{13}+  +   \eta_2 \eta_3 s^{23}  + \eta_1 \eta_2 \eta_3 f_-^{123} .  \label{eq:n=3phi}
\end{split}
}
This superfield includes the positive helicity gluon, so to get the CPT complete multiplet we construct the $\Psi$ superfield by integrating $\eta_4$ out
\Eq{ 
\begin{split}
\Psi^{\mathcal{N}  =  3} &= \int \eta_4 \Phi^{\mathcal{N}=4} \\
&= f^{(4)}_+  -  \eta_1  s^{1(4)} - \eta_2  s^{2(4)} -  \eta_3  s^{3(4)}  +  \eta_1 \eta_2  f_-^{12(4)} \\
&+  \eta_1 \eta_3  f_-^{13(4)} + \eta_2 \eta_3 f_-^{23(4)}  - \eta_1 \eta_2 \eta_3  g_-^{123(4)}, \label{eq:n=3psi}
\end{split}
}
where we have introduced the indices in parenthesis, which are ones that have been integrated out. The superfields \eqref{eq:n=3phi} and \eqref{eq:n=3psi} are related by CPT conjugation, so combined they constitute the full field content of the $\mathcal{N}=3$ SYM multiplet (which is identical to the field content of the $\mathcal{N}=4$ SYM multiplet). We can remove $\eta_3$ (set it to zero in $\Phi$ and integrate it out in $\Psi$) to get the $\mathcal{N}=2$ SYM superfields and so on. 

\begin{figure}[t!]
\centering 
{\LARGE
\eq{
 \vcenter{\hbox{\includegraphics[height=1.9cm]{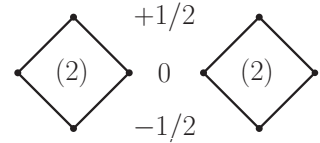}}} 
}}
\captionsetup{justification=raggedright, margin = 15pt} 
\caption{This diamond diagram, representing the superfields \eqref{eq:chiralfromsym}. Each diamond represents a superfield in \eqref{eq:chiralfromsym}. The top of the diamonds denotes the $+1/2$ helicity state, the middle is the two $0$ helicity states and the bottom is the $-1/2$ helicity state. The numbers in parentheses denotes the number of states at a certain helicity and no number means that there is one state the that specific helicity. The lines in the diamonds denote the relation between states through supersymmetry.} \label{fig:chiralfromsym}
\end{figure}

\begin{figure}[t!]
\centering 
{\LARGE
\eq{
 \vcenter{\hbox{\includegraphics[height=1.9cm]{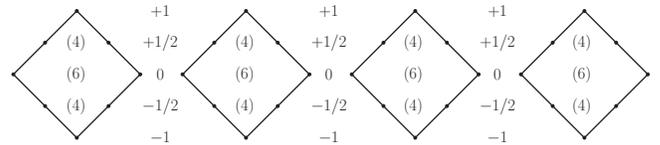}}} 
}}
\captionsetup{justification=raggedright, margin = 20pt} 
\caption{This diamond diagram, representing the fields \eqref{eq:vectorfromsugra}, is divided into four diamonds representing each of the four vector superfields of equation \eqref{eq:vectorfromsugra}.} \label{fig:vectorfromsugra}
\end{figure}

We will now introduce a procedure for reducing supersymmetry, which will give us a way to construct chiral multiplets from the SYM multiplet. If we truncate the $\mathcal{N}=4$ SYM superfield from both the top and the bottom, removing the vector degrees of freedom, we end up with the desired chiral multiplet. The only way we can achieve this construction is if we integrate out one of the Grassmann variables in the set $(\eta_1, \eta_4)$ and set the other to zero. This yields two superfields, namely one where we integrate out $\eta_1$ and then set $\eta_4$ to zero and vice versa
\Eq{
\begin{split} \label{eq:chiralfromsym}
\chi^{\mathcal{N}=2}_1 &=  \left. \int d\eta_1 \Phi^{\mathcal{N}=4} \right|_{\eta_4\rightarrow 0} ,    \\
\chi^{\mathcal{N}=2}_2 &=  \left. \int d\eta_4 \Phi^{\mathcal{N}=4} \right|_{\eta_1 \rightarrow 0} .
\end{split}
}
See the Appendix for the explicit expressions for these superfields. The superfields \eqref{eq:chiralfromsym} can also be represented in terms of diamond diagrams \cite{kltmap}, see Figure \ref{fig:chiralfromsym}.
Since our goal is to reduce the KLT relations \eqref{eq:N=8superklt} to relations for squaring chiral amplitudes, we also need to apply the above reduction procedure to the SUGRA superfield \eqref{eq:n=8superfield}. In the KLT relations, the $SU(8)_R$ indices on the SUGRA side are related to the $SU(4)_R$ indices by taking the union of the $SU(4)_R$ indices. If we use the supersymmetry reduction procedure on both SYM multiplets in equation \eqref{eq:N=8superklt} then on the SUGRA side we get four different ways (depending on which of the reduction methods in \eqref{eq:chiralfromsym} is considered on each SYM multiplet) of reducing the $\mathcal{N}=8$ SUGRA superfield into vector superfields 
\Eq{
\begin{split} \label{eq:vectorfromsugra}
\nu^{\mathcal{N}=4}_1 &= \left. \int d\eta_1 d\eta_5 \Phi^{\mathcal{N}_G = 8} \right|_{\eta_4, \eta_8 \rightarrow 0} ,  \\
\nu^{\mathcal{N}=4}_2 &= \left. \int d\eta_4 d\eta_5 \Phi^{\mathcal{N}_G = 8} \right|_{\eta_1, \eta_8 \rightarrow 0} , \\
\nu^{\mathcal{N}=4}_3 &= \left. \int d\eta_1 d\eta_8 \Phi^{\mathcal{N}_G = 8} \right|_{\eta_4, \eta_5 \rightarrow 0} ,  \\
\nu^{\mathcal{N}=4}_4 &= \left. \int d\eta_4 d\eta_8 \Phi^{\mathcal{N}_G = 8} \right|_{\eta_1, \eta_5 \rightarrow 0} .
\end{split}  
}
See the Appendix for the explicit expressions of these superfields. We can also represent the superfields of equation \eqref{eq:vectorfromsugra} in terms of diamond diagrams, see Figure \ref{fig:vectorfromsugra}.
Next up, we will apply these supersymmetry reduction methods to the full KLT relations \eqref{eq:N=8superklt} and investigate the properties of the resulting relations.


\section{Chiral KLT relations and chiral squaring  \label{sec:chiralklt}}

Applying the procedure of supersymmetry reduction, from the previous section, to the two copies of SYM multiplets in equation \eqref{eq:N=8superklt} and simultaneously to SUGRA multiplet, the resulting equation will relate chiral amplitudes and vector amplitudes. Hence we have constructed explicit chiral KLT relations, which was our initial goal of this paper. These relations take the explicit form
\begin{widetext}
\Eq{ \label{eq:chiralkltrelation}
\begin{split}
&\mathcal{M}_n^{\mathcal{N} = 4} \left[ (\nu_1)^{\mathcal{N} = 4}_{i_1, \ldots, i_{m_1}}, (\nu_2)^{\mathcal{N} = 4 }_{j_1, \ldots, j_{m_2}}, (\nu_3)^{\mathcal{N} = 4}_{k_1, \ldots, k_{m_3}}, (\nu_4)^{\mathcal{N} = 4 }_{l_1, \ldots, l_{m_4}} \right] = \\
&\sum_{\gamma, \beta \in S_{n-3}} \mathcal{A}_n^{\mathcal{N} = 2 } \left[(\chi_1)^{\mathcal{N}=  2}_{i_1, \ldots, i_{m_1}; k_1, \ldots, k_{m_3}}, (\chi_2)^{\mathcal{N} = 2 }_{j_1, \ldots, j_{m_2}; l_1, \ldots, l_{m_4}} \right]\mathcal{S}[\gamma| \beta]_{p_1}  \tilde{\mathcal{A}}_n^{ \tilde{\mathcal{N}} = 2 } \left[ (\chi_1)^{\tilde{\mathcal{N}} = 2 }_{i_1, \ldots, i_{m_1}; j_1, \ldots, j_{m_2}}, (\chi_2)^{ \tilde{\mathcal{N}} = 2 }_{k_1, \ldots, k_{m_3}; l_1, \ldots, l_{m_4}}\right],
\end{split}
}
\end{widetext}
where $m_1 + m_2 + m_3 + m_4 = n$ and $i_1, \ldots, i_{m_1}$ are in the $\nu_1$ representation, legs $j_1, \ldots, j_{m_2}$ are in the $\nu_2$ representation, legs $k_1, \ldots, k_{m_3}$ are in the $\nu_3$ representation and legs $l_1, \ldots, l_{m_4}$ are in the $\nu_4$ representation. See the Appendix for the exact superfield expressions. Equation \eqref{eq:chiralkltrelation} makes the 'squaring' of the chiral amplitudes manifest, resulting in vector amplitudes, and hence realizes the notion of chiral squaring. This is also apparent from the superfield relations
\Eq{ \label{eq:chiralsquaring}
\begin{split}
\nu_1 &= \chi_1 \otimes \tilde{\chi}_1 ,\\
\nu_2 &= \chi_2 \otimes \tilde{\chi}_1, \\
\nu_3 &= \chi_1 \otimes \tilde{\chi}_2 ,\\
\nu_4 &= \chi_2 \otimes \tilde{\chi}_2.
\end{split}
}
These chiral squaring relations can also be represented in terms of diamond diagrams,
\Eq{
\begin{split}
&\left[ \vcenter{\hbox{\includegraphics[height=1.35cm]{chiralfromsym.eps}}} \right]  \otimes \left[ \vcenter{\hbox{\includegraphics[height=1.35cm]{chiralfromsym.eps}}}\right] \\
&=  \left[ \vcenter{\hbox{\includegraphics[height=1.44cm]{vectorfromsugra.eps}}} \right].
\end{split}
}
The ordering of legs on the chiral side of \eqref{eq:chiralkltrelation} is the same as for the SYM amplitudes in equation \eqref{eq:N=8superklt} and the $\mathcal{S}$-kernel is given in equation \eqref{eq:s-kernel}. The field content of \eqref{eq:chiralkltrelation} is given in \eqref{eq:chiralfromsym} for the chiral superfields and \eqref{eq:vectorfromsugra} for the vector superfields. We note that there are no restriction on how many times each superfield has to appear, as each superfield on both the chiral side and the vector side contains the same Grassmann variables.

Now that we have the chiral KLT relations \eqref{eq:chiralkltrelation}, there are some peculiarities that we need to address regarding the amplitudes in question. On the chiral side we have amplitudes similar to Yukawa theory amplitudes. However these amplitudes are color-ordered (since they originate from SYM amplitudes) and hence we expect vanishing relations like KK and BCJ relations \cite{kk, bcj, kkbcjproof} for these chiral amplitudes. BCJ relations should be realized through the $\mathcal{S}$-kernel, as it can be seen as a generator of BCJ relations. Specifically for each $\gamma_{2,n-1}$ permutation in \eqref{eq:chiralkltrelation}, we can write a general combination of BCJ relations \cite{kltfieldtheory} as
\Eq{ \label{eq:chiralbcj}
\sum_{ \beta \in S_{n-3}} \mathcal{S}[\gamma_{2, n-1} | \beta_{2, n-1} ]_{p_1}  \mathcal{A}_n^{ \mathcal{N} = 2 } \left[ 1, \beta_{2, n-1}, n \right] = 0.
}
Another way of constructing vanishing relations for the chiral amplitudes is to use, that superamplitudes can be decomposed into an MHV part, an NMHV part and so on (see equation \eqref{eq:N=4SYMsuperamplitudeexpansion} for the $\mathcal{N}=4$ SYM superamplitude expansion). For the chiral amplitudes, we expect an expansion like
\Eq{ \label{eq:n=2chiralsuperamplitudeexpansion}
\begin{split}
\mathcal{A}_n^{\mathcal{N}=2} &= \sum A_n^\text{MHV} (\eta)^4 + \sum A_n^\text{NMHV} (\eta)^{6}  \\
&+ \cdots + \sum A_n^{\overline{\text{MHV}}} (\eta)^{2n-4}
\end{split}
}
Going back to equation \eqref{eq:chiralkltrelation} we can violate $R$-symmetry on the vector side by picking out a component amplitude in the un-tilded sector on the chiral side as a N$^k$MHV amplitude while picking an N$^{k'}$MHV amplitude in the tilded sector. If $k \ne k'$, then all Grassmann variables on the vector side would not appear the same number of times, and hence $R$-symmetry is violated and the vector amplitude vanishes. However each amplitude on chiral side are still $R$-invariant, so these are indivdually non-vanshing. Hence we get a nontrivial vanishing relation
\Eq{ \label{eq:chiralvanishing}
\begin{split}
&\sum_{\gamma, \beta \in S_{n-3}} A_n^{N^k MHV} \mathcal{S}[\gamma| \beta]_{p_1}  A_n^{ N^{k'} MHV}  = 0,
\end{split} 
}
when $k \ne k'$. 

Something we haven't touched upon yet is the interpretation of the field content on the the vector side of \eqref{eq:chiralkltrelation}. These amplitudes come from an on-shell vector multiplet, but what are the dynamics and interactions for this multiplet? The amplitudes do not describe any form of SYM type multiplets as the amplitudes are fully symmetric with respect to momentum permutations. The amplitudes have restrictions through equation \eqref{eq:chiralkltrelation} such that certain amplitudes, which would be nonzero in SYM theory, are vanishing due to restrictions on fermionic configurations on the chiral side of \eqref{eq:chiralkltrelation}. There exists no 3-point MHV/anti-MHV amplitude, since a 3-point pure vector amplitude would factorize into two 3-point pure fermion amplitudes, which both vanish. As it turns out, the vector states are actually the gravitational equivalent of a photon, called a graviphoton, and it acts in many ways like a $U(1)$-like gauge boson \cite{antigravity1, antigravity2, graviphotonL}. In terms of phenomenology, the graviphoton couples to the mass currents and acts as a repulsive force under graviphoton exchange between two (anti)matter particles hence leading to an anti-gravitational force.


\section{Chiral KLT relations: Explicit checks \label{sec:explicit}}

In this section we will do some explicit checks of the chiral KLT relation \eqref{eq:chiralkltrelation}. The relation should be manifest, as it has been derived from an identity. Therefore we make these checks to show how the relation works in a practical sense. 

To make things very simple, we only consider particles in the $\nu_1$ representation on the vector side (see equation \eqref{eq:vectorfromsugra} and the Appendix). We can justify this choice, without loss of generality, since there is no inherent difference between the four vector multiplets. Hence the external superfields on the chiral side will be in the $\chi_1$ representation for both sectors (see equation \eqref{eq:chiralsquaring}). The resulting simplified chiral KLT relation is then
\Eq{ \label{eq:chiralkltrelationsimplified}
\begin{split}
\mathcal{M}_n^{\mathcal{N} = 4} \left[ (\nu_1)_i \right] = &\sum_{\gamma, \beta \in S_{n-3}} \mathcal{A}_n^{\mathcal{N} = 2 } \left[(\chi_1)_i \right]  \\
&\times \mathcal{S}[\gamma| \beta]_{p_1} \times   \tilde{\mathcal{A}}_n^{ \tilde{\mathcal{N}} = 2 } \left[ (\chi_1)_i \right].
\end{split}
}
We will do explicit checks of 3- and 4-point relations. However, this is just to show how the relation \eqref{eq:chiralkltrelation} works in practice. The chiral KLT relation is derived from an identity (the full KLT relation \eqref{eq:N=8superklt}), so we expect no unusual behaivior in these checks. The 3-point chiral KLT relations are simply
\Eq{ \label{eq:chiralkltrelation3pt}
\mathcal{M}_3^{\mathcal{N} = 4} \left[ 1, 2, 3\right] &= \mathcal{A}_3^{\mathcal{N} = 2 } \left[1,2,3\right]    \tilde{\mathcal{A}}_3^{ \tilde{\mathcal{N}} = 2 } \left[ 1, 2, 3  \right] ,
}
with the definition $\mathcal{S}[\emptyset | \emptyset]_{p_1} =1$. The 4-point relations are slightly more complicated, taking the form
\Eq{ \label{eq:chiralkltrelation4pt}
\begin{split}
&\mathcal{M}_4^{\mathcal{N} = 4} \left[ 1, 2, 3, 4\right] \\
&= s_{12} \mathcal{A}_4^{\mathcal{N} = 2 } \left[3 , 4, 2, 1 \right]    \tilde{\mathcal{A}}_4^{ \tilde{\mathcal{N}} = 2 } \left[ 1, 2, 3, 4 \right].
\end{split}
} 
We pick out any desired particles for the external states by applying derivatives, similarly to picking out specific states from the $\mathcal{N}=4$ SYM superfield in Table \ref{table:n=4superfield}. Let us first consider any sensible 3-point amplitude that are local interactions. Let us consider, on the vector side, a 3-point amplitude with two fermions and a vector particle \cite{graviphotonL}. We can apply derivatives on the lefthand side of equation \eqref{eq:chiralkltrelation3pt}
\Eq{
\begin{split}
 \mathcal{M}_3^{\mathcal{N} = 4} \left[ 1, 2, 3\right] \rightarrow M_3 \left[ f^{2+ }_1 f^{2 + }_2 3^-  \right] .
\end{split}
}
The notation $3^-$ means that we have a vector particle on leg 3 with negative helicity and $f^{2+}_1$ means a fermion on leg 1 with positive helicity and $R$-index 2. On the chiral side of \eqref{eq:chiralkltrelation} we apply the same derivatives and get
\Eq{
 A_3 [ \phi^2_1 \phi^2_2 f^-_3 ] A_3 [f^+_1 f^+_2 f^-_3]  = 0.
}
This product is zero since both amplitudes are vanishing. Amplitudes with an odd number of fermions are not Lorentz invariant and hence must vanish. However we can consider another chiral KLT relation, which is nontrivial,
\Eq{  \label{eq:chiralklt3ptyukawa}
 M_3 \left[ f^{2 +}_1 f^{6 + }_2 3^+  \right] = A_3 [ \phi^2_1 f^+_2 f^+_3 ] A_3 [f^+_1 \phi_2^6 f^+_3].
}
There are four different versions of this relation (depending on the choice of $SU(4)_R$ indices on the fermions on the vector side). We can calculate any of these 3-point amplitudes using Little group scaling and spinor helicity conventions from \cite{elvang}. For the vector amplitudes we get $M_3 \left[ f^{+}_1 f^{+ }_2 3^+  \right] = [13] [23]$. For the chiral amplitudes we get $A_3 [ \phi_1 f^+_2 f^+_3 ] = [23]$ and $A_3 [f^+_1 \phi_2 f^+_3] = [13]$. Hence the relations \eqref{eq:chiralklt3ptyukawa} work out fine. For the all minus helicity version of equation \eqref{eq:chiralklt3ptyukawa}, we switch out square brackets with angle brackets. We can also consider a non-minimal coupling \cite{graviphotonL} between two graviphotons and a scalar
\Eq{
M_3 \left[ 1^- 2^- \phi_3^{26}  \right] &= A_3 [ f^-_1 f^-_2 \phi_3^6 ] A_3 [f^-_1 f^-_2 \phi_3^6 ] .
}
Again using Little group scaling, we find the vector amplitude
\Eq{ \label{eq:graviphoton3pt}
M_3 \left[ 1^- 2^- \phi_3^{26}  \right]  = \braket{12}^2,
}
and the chiral amplitude
\Eq{ \label{eq:yukawa3pt}
 A_3 [ f^-_1 f^-_2 \phi_3^6 ] = \braket{12} = A_3 [f^-_1 f^-_2 \phi_3^6 ] ,
}
which means that the explicit relation works out.

Having determined some of the 3-point relations we move on to do some more nontrivial checks 4-point (see equation \eqref{eq:chiralkltrelation4pt}). Consider the exchange of a graviphoton between two fermions
\Eq{ \label{eq:chiralklt4ptfermion}
\begin{split}
&M_4 \left[ f^{2 +}_1 f^{6 + }_2 f^{ 267 -}_3 f^{236-}_4  \right] \\
&=  s_{12} A_4 \left[ \phi_1^2 f_2^+ \phi^2_3 f_4^- \right] A_4 \left[ f_1^+ \phi^6_2 f_3^- \phi^6_4 \right] 
\end{split}
}
Both sides are constructible using the BCFW recursion relations \cite{elvang} from the amplitudes in \eqref{eq:chiralklt3ptyukawa}. The resulting amplitude on the vector side is quite simple: $M_4 \left[ f^{2 +}_1 f^{6 + }_2 f^{ 267 -}_3 f^{236-}_4  \right] = s_{12} \frac{ \braket{34}^2}{\braket{12}^2}$. On the chiral side we get amplitudes $A_4 \left[ \phi_1^2 f_2^+ \phi^2_3 f_4^- \right]  = A_4 \left[ f_1^+ \phi^6_2 f_3^- \phi^6_4 \right]  = \frac{\braket{34}}{\braket{12}}$. Hence equation \eqref{eq:chiralklt4ptfermion} works out perfectly. 

Consider now a case where the external particles on the vector side are only graviphotons. We take the 4-point chiral KLT relation \eqref{eq:chiralkltrelation4pt} with two negative helicity graviphotons on legs 1 and 2 and two positive helicity graviphotons on legs 3 and 4
\Eq{ \label{eq:chiralklt4ptvectorscat}
\begin{split}
&M_4 \left[ 1^- 2^- 3^+ 4^+ \right]  \\
&= s_{12} A_4 [ f^-_1 f^-_2 f^+_3 f^+_4 ] A_4 [ f^-_1 f^-_2 f^+_4 f^+_3 ] 
\end{split}
}
We construct amplitudes from the 3-point amplitudes in equations \eqref{eq:graviphoton3pt} and \eqref{eq:yukawa3pt}. In the BCFW recursion there are now internal scalars to be summed over. For the amplitudes \eqref{eq:graviphoton3pt}, we have six choices of scalars from the vector multiplet. However only four choices of scalars in the 3-point amplitude yields a nonzero amplitude \eqref{eq:graviphoton3pt}. For example 
\Eq{
M_3 \left[ 1^- 2^- \phi_3^{23}  \right] &= A_3 [ f^-_1 f^-_2 f^+_3 ] A_3 [f^-_1 f^-_2 f^-_3] = 0.
}
Similar vanishing happens when picking the scalar $\phi^{67}$. There is a similar non-trivial counting for constructing the chiral amplitudes in equation \eqref{eq:chiralklt4ptvectorscat}, since there are two possible BCFW-internal scalars in each chiral multiplet. Taking this into account, we get a counting factor for each amplitude we calculate. The vector amplitude in question ends up taking the form
\Eq{
M_4 \left[ 1^- 2^- 3^+ 4^+ \right] = 4 \frac{\braket{12}^2 [34]}{\braket{34}},
}
where the factor of 4 count the four scalars that can run internally in the BCFW relations. A similar counting factor shows up, in the chiral amplitude, due to the two scalars in the multiplet
\Eq{
A_4 [ f^-_1 f^-_2 f^+_3 f^+_4 ] &= 2 \frac{\braket{12}}{\braket{34}},
}
It can now easily be checked that the relation \eqref{eq:chiralklt4ptvectorscat} holds.


\section{Conclusion}

In this paper we have shown how to perform KLT-squaring of chiral multiplets. In order to do so we have developed a new procedure for reducing supersymmetry in a maximally supersymmetric multiplet. In Section \ref{sec:susy} we found a procedure, akin to the one used in \cite{kltmap}, for reducing the $\mathcal{N} =4$ SYM multiplet to two $\mathcal{N} =2$ chiral multiplets and similarly $\mathcal{N} =8$ SUGRA into four copies of $\mathcal{N} =4$ vector multiplets.

Using this reduction procedure with KLT relations, described in Section \ref{sec:klt}, we derive the chiral KLT relations \eqref{eq:chiralkltrelation}, where chiral squaring is realized. Similar ideas have been suggested in \cite{nagy}, where a more algebraic and group theoretic approach was taken. In this paper we find chiral squaring resulting directly from the KLT relations.

The chiral amplitudes in \eqref{eq:chiralkltrelation} must satisfy KK and BCJ relations since they replace the SYM amplitudes in equation \eqref{eq:N=8superklt}. The $\mathcal{S}$-kernel generates BCJ relations for the chiral amplitudes like for SYM amplitudes. From the chiral KLT relations we also find vanishing relations from violating $R$-symmetry on the vector side. Finally we have shown how the chiral KLT relations work in simple examples.
\vspace{1mm}


\begin{acknowledgments}
We would like to thank N. E. J. Bjerrum Bohr and P. H. Damgaard, as well as C. Liu, for helpful discussions.
\end{acknowledgments}


\appendix*

\section{Appendix: chiral KLT relation superfields: explicit expressions}

Here we present the explicit superfield expressions used in equation \eqref{eq:chiralkltrelation} and shown schematically in equation \eqref{eq:chiralfromsym} and equation \eqref{eq:vectorfromsugra}. The superfield are calculated by starting with the maximally supersymmetric superfields (either equation \eqref{eq:n=4superfield} or equation \eqref{eq:n=8superfield}) and then removing Grassmann variables from the superfield. For the chiral superfields we start from equation \eqref{eq:n=4superfield} and get the following expressions
\Eq{
\begin{split}
\chi^{\mathcal{N}=2}_1 &=   f^{(1)}_+  +  \eta_2 s^{(1)2} +   \eta_3 s^{(1)3}+  + \eta_2 \eta_3 f_-^{(1)23},    \\
\chi^{\mathcal{N}=2}_2 &=  f^{(4)}_+   -  \eta_2  s^{2(4)}-   \eta_3 s^{3(4)}  + \eta_2 \eta_3  f_-^{23(4)} .
\end{split}
}
For the vector superfields we start from equation \eqref{eq:n=8superfield} and arrive at the following expressions
\begin{widetext}
\Eq{
\begin{split}
\nu^{\mathcal{N}=4}_1 &=  -  \nu^{(15)}_+ - \eta_A \chi^{A(15)}_+ - \frac{1}{2!} \eta_A \eta_B  \phi^{AB(15)} - \frac{1}{3!} \eta_A \eta_B \eta_C  \chi_-^{ABC(15)} - \eta_2 \eta_3 \eta_6 \eta_7  \nu_-^{2367(15)} , \\
\nu^{\mathcal{N}=4}_2 &=  -   \nu^{(45)}_+ -  \eta_A  \chi^{A(45)}_+ - \frac{1}{2!} \eta_A \eta_B \phi^{AB(45)} - \frac{1}{3!} \eta_A \eta_B \eta_C \chi_-^{ABC(45)} -  \eta_2 \eta_3 \eta_6 \eta_7 \nu_-^{2367(45)},  \\
\nu^{\mathcal{N}=4}_3 &=  -   \nu^{(18)}_+ -  \eta_A  \chi^{A(18)}_+ - \frac{1}{2!} \eta_A \eta_B \phi^{AB(18)} - \frac{1}{3!} \eta_A \eta_B \eta_C \chi_-^{ABC(18)} -  \eta_2 \eta_3 \eta_6 \eta_7 \nu_-^{2367(18)},  \\
\nu^{\mathcal{N}=4}_4 &=  -   \nu^{(48)}_+ -  \eta_A  \chi^{A(48)}_+ - \frac{1}{2!} \eta_A \eta_B \phi^{AB(48)} - \frac{1}{3!} \eta_A \eta_B \eta_C \chi_-^{ABC(48)} -  \eta_2 \eta_3 \eta_6 \eta_7 \nu_-^{2367(48)},
\end{split}  
}
\end{widetext}
where $A,B,C = 2,3,6,7$ are $SU(4)_R$ indices.


\end{document}